\documentclass[12pt]{article}
\usepackage{amsmath,url}
\usepackage{graphicx}
\usepackage{cite}
\input{epsf} 
\providecommand{\href}[2]{#2}

\newcommand\as{\alpha_{\mathrm{S}}} 
\newcommand\f[2]{\frac{#1}{#2}} 
 
\def\to{\rightarrow}

\begin{document} 
\begin{titlepage}
\renewcommand{\thefootnote}{\fnsymbol{footnote}}
\begin{flushright}
ZU-TH 19/14\\
MITP/14-033
\end{flushright}
\vspace*{2cm}

\begin{center}
{\Large \bf $ZZ$ production at hadron colliders in NNLO QCD}
\end{center}

\par \vspace{2mm}
\begin{center}
{\sc F.~Cascioli$^{(a)}$, T.~Gehrmann$^{(a)}$, M.~Grazzini$^{(a)}$\footnote{On leave of absence from INFN, Sezione di Firenze, Sesto Fiorentino, Florence, Italy.}, S.~Kallweit$^{(a)}$, P.~Maierh{\"o}fer$^{(a)}$, A.~von~Manteuffel$^{(b)}$, S.~Pozzorini$^{(a)}$, D.~Rathlev$^{(a)}$, L.~Tancredi$^{(a)}$} and {\sc E.~Weihs$^{(a)}$}
\vspace{5mm}

$^{(a)}$Physik-Institut, Universit\"at Z\"urich, CH-8057 Z\"urich, Switzerland 

$^{(b)}$PRISMA Cluster of Excellence, Institute of Physics,
Johannes Gutenberg University,\\ D-55099 Mainz, Germany

\vspace{5mm}

\end{center}

\par \vspace{2mm}
\begin{center} {\large \bf Abstract} \end{center}
\begin{quote}
\pretolerance 10000

We report on the first calculation of next-to-next-to-leading order (NNLO) QCD corrections to the inclusive production of $Z$-boson pairs at hadron colliders. Numerical results are presented for $pp$ collisions
with centre-of-mass energy ($\sqrt{s}$) ranging from 7 to 14 TeV.
The NNLO corrections increase the NLO result by an amount
varying from 11\% to 17\% as $\sqrt{s}$ goes from 7 to 14 TeV. The loop-induced gluon fusion contribution provides about $60\%$ of the total NNLO effect.
When going from NLO to NNLO the scale uncertainties do not decrease and remain at the $\pm 3\%$ level.

\end{quote}

\vspace*{\fill}
\begin{flushleft}
May 2014

\end{flushleft}
\end{titlepage}

\setcounter{footnote}{1}
\renewcommand{\thefootnote}{\fnsymbol{footnote}}

The production of vector-boson pairs is a crucial process
for physics studies within and beyond the Standard Model~(SM).
In particular the production of $Z$-boson pairs is an irreducible background
for Higgs boson production and new-physics searches.
Various measurements of $ZZ$ hadroproduction have been carried out at the Tevatron and the LHC
(for some recent results see Refs.~\cite{CDF:2011ab,Abazov:2012cj,Aad:2012awa,Chatrchyan:2012sga,ATLAS-CONF-2013-020,CMS-PAS-SMP-13-005}).

The theoretical efforts for a precise prediction of $ZZ$ production in the Standard Model
started more than 20 years ago, with the first NLO QCD calculations
\cite{Ohnemus:1990za,Mele:1990bq} with stable $Z$ bosons.
The leptonic decays of the $Z$ bosons
were then added, initially neglecting spin correlations in the virtual 
contributions \cite{Ohnemus:1994ff}.
The computation of the relevant one-loop helicity amplitudes \cite{Dixon:1998py} allowed
complete NLO calculations \cite{Campbell:1999ah,Dixon:1999di} including
spin correlations and off-shell effects.
The loop-induced gluon fusion contribution, which is formally next-to-next-to-leading order (NNLO),
has been computed in Refs.~\cite{Glover:1988rg,Dicus:1987dj}. 
The corresponding leptonic decays have been
included in Refs.~\cite{Matsuura:1991pj,Zecher:1994kb,Binoth:2008pr}.
Since the gluon-induced contribution is enhanced by the gluon luminosity, it is often assumed to
provide the bulk of the NNLO corrections.
NLO predictions for $ZZ$ production including the gluon-induced contribution,
the leptonic decay with spin correlations and off-shell effects have
been presented in Ref.~\cite{Campbell:2011bn}.
The NLO QCD corrections to on-shell $ZZ+{\rm jet}$ production have been discussed in Refs.~\cite{Binoth:2009wk,Binoth:2010ra}, and the electroweak (EW) corrections to $ZZ$ production have been computed in Refs.~\cite{Bierweiler:2013dja,Baglio:2013toa}.

In this Letter we report on the first calculation of the inclusive production 
of on-shell $Z$-boson pairs at hadron colliders in NNLO QCD.

The NNLO computation requires the evaluation of the tree-level
scattering amplitudes with two additional (unresolved) partons, of the one-loop amplitudes with one additional parton, and of the one-loop-squared and
two-loop corrections to the Born subprocess $q{\bar q}\to ZZ$.
All the relevant tree and one-loop matrix elements are automatically generated with {\sc OpenLoops}~\cite{Cascioli:2011va}, which implements
a fast numerical recursion 
for the calculation of NLO scattering amplitudes within the SM. For the numerically stable evaluation of tensor integrals we rely on the {\sc Collier} library~\cite{collier}, which is based on the Denner--Dittmaier reduction techniques~\cite{Denner:2002ii,Denner:2005nn} and the scalar integrals of~\cite{Denner:2010tr}.
The loop-induced gluon fusion contribution is also obtained with {\sc OpenLoops}, including five light-quark flavors and massive
top-quark loops\footnote{Consistently with the inclusion of five active flavors, the renormalisation of the QCD coupling $\as$ is performed in the so-called decoupling scheme, where top-quark loops are subtracted at zero momentum transfer. In this scheme, 
the $q\bar q\to ZZ g$, $q g\to ZZ q$ and $\bar q g \to ZZ\bar q$ channels receive top-quark contributions only via ultraviolet-finite box diagrams, while the top-quark contributions to the gluon-field and $\as$ counterterms cancel against each other.}.
The SM Higgs boson contribution is also considered.
Following the recent computation of the relevant two-loop master integrals \cite{Gehrmann:2013cxs,Henn:2014lfa,Gehrmann:2014bfa,Caola:2014lpa}
the last missing contribution,
the genuine two-loop correction to the $ZZ$ amplitude,
has been computed by some of us, and will be reported elsewhere \cite{inprep}.
In the two-loop correction, contributions involving a top-quark loop
are neglected. For the numerical evaluation of the multiple
polylogarithms in the two-loop expressions we employ the
implementation \cite{Vollinga:2004sn} in the {\sc GiNaC} \cite{Bauer:2000cp}
library.

The implementation of the various scattering amplitudes in a complete NNLO calculation is
a highly non-trivial task due to the presence of infrared~(IR) singularities at
intermediate stages of the calculation that prevent a straightforward application of numerical techniques.
To handle and cancel these singularities at NNLO
we employ the $q_T$ subtraction method \cite{Catani:2007vq}.
This approach applies to the
production of a colourless high-mass system $F$
in generic hadron collisions and has been used for the computation of
NNLO corrections to several hadronic processes \cite{Catani:2007vq,Catani:2009sm,Ferrera:2011bk,Catani:2011qz,Grazzini:2013bna}.
According to the $q_T$ subtraction method \cite{Catani:2007vq}, the $pp\to F+X$ cross section at NNLO
can be written as
\begin{equation}
\label{main}
d{\sigma}^{F}_{NNLO}={\cal H}^{F}_{NNLO}\otimes d{\sigma}^{F}_{LO}
+\left[ d{\sigma}^{F+{\rm jet}}_{NLO}-
d{\sigma}^{CT}_{NLO}\right]\;\; ,
\end{equation}
where $d{\sigma}^{F+{\rm jet}}_{NLO}$ is the cross section for the
inclusive production of the system $F$ plus one jet at NLO accuracy, and can be evaluated with
any available version of the NLO subtraction formalism.
When the transverse momentum $q_T$ of the colourless system $F$ is non-vanishing,
$d{\sigma}^{F+{\rm jet}}_{NLO}$ is the sole contribution to the NNLO cross section.
The IR subtraction counterterm $d{\sigma}^{CT}_{NLO}$ in Eq.~(\ref{main}) has the purpose of cancelling
the singularity developed by $d{\sigma}^{F+{\rm jet}}_{NLO}$ as $q_T\to 0$ and
is obtained from the resummation of the logarithmically-enhanced
contributions to $q_T$ distributions \cite{Bozzi:2005wk}.  
The function ${\cal H}^{F}_{NNLO}$, which also compensates for the subtraction
of $d{\sigma}^{CT}_{NLO}$,
corresponds to the NNLO truncation of the process-dependent perturbative function
\begin{equation}
{\cal H}^{F}=1+\f{\as}{\pi}\,
{\cal H}^{F(1)}+\left(\f{\as}{\pi}\right)^2
{\cal H}^{F(2)}+ \dots \;\;.
\end{equation}
The NLO calculation  of $d{\sigma}^{F}$ 
requires the knowledge
of ${\cal H}^{F(1)}$, and the NNLO calculation also requires ${\cal H}^{F(2)}$.

The general structure of ${\cal H}^{F(1)}$
is known \cite{deFlorian:2001zd}: 
${\cal H}^{F(1)}$ is obtained from the process-dependent scattering
amplitudes by using a process-independent relation.
Exploiting the explicit results of ${\cal H}^{F(2)}$ for Higgs
\cite{Catani:2011kr} and vector-boson \cite{Catani:2012qa} 
production,
the process-independent relation of 
Ref.~\cite{deFlorian:2001zd} has been extended to the calculation of the NNLO coefficient 
${\cal H}^{F(2)}$ \cite{Catani:2013tia}. Such results have been confirmed
with a fully independent calculation of the relevant coefficients in the framework of Soft-Collinear Effective Theory~(SCET) \cite{Gehrmann:2012ze,Gehrmann:2014yya}.
We have performed our NNLO calculation for $ZZ$ production
according to Eq.~(\ref{main}), starting from a computation
of the $d{\sigma}^{ZZ+{\rm jet}}_{NLO}$ cross section with the dipole-subtraction method \cite{Catani:1996jh,Catani:1996vz}.
The numerical calculation employs the generic Monte Carlo program that was
developed for Ref.~\cite{Grazzini:2013bna}.
Although the $q_T$ subtraction method and our implementation are suitable to perform a fully exclusive computation of $ZZ$ production including the leptonic decays and the corresponding spin correlations, in this Letter we restrict ourselves to the inclusive production of on-shell $Z$ bosons.

\begin{figure}[ht]
\begin{center}
\begin{tabular}{cc}
\includegraphics[width=0.47\textwidth,angle=90]{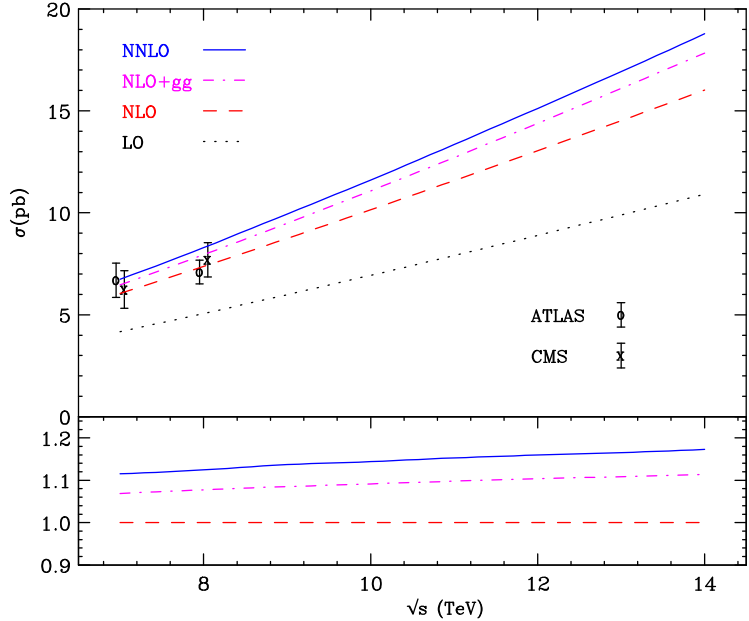}\\
\end{tabular}
\end{center}
\caption{{\em $ZZ$ cross section at LO (dots), NLO (dashes), NLO+gg (dot dashes) and NNLO (solid) as a function of $\sqrt{s}$. The ATLAS and CMS experimental results at $\sqrt{s}=7$ TeV and $\sqrt{s}=8$ TeV are also shown for comparison \cite{Aad:2012awa,Chatrchyan:2012sga,ATLAS-CONF-2013-020,CMS-PAS-SMP-13-005}. The lower panel shows the NNLO and NLO+gg results normalized to the NLO prediction.}}
\end{figure}

We consider $pp$ collisions with $\sqrt{s}$ ranging from 7 to 14 TeV.
As for the EW couplings, we use the so-called $G_\mu$ scheme,
where the input parameters are $G_F$, $m_W$, $m_Z$. In particular we 
use the values
$G_F = 1.16639\times 10^{-5}$~GeV$^{-2}$, $m_W=80.399$ GeV,
$m_Z = 91.1876$~GeV.
The top mass $m_t=173.2$ GeV and the Higgs mass $m_H=125$ GeV only enter through the loop-induced gluon fusion contribution\footnote{Since we consider the production of on-shell $Z$ bosons, the Higgs contribution is strongly suppressed,
and provides only about $1\%$ to the loop-induced $gg\to ZZ$ cross section.}.
We use the MSTW 2008 \cite{Martin:2009iq} sets of parton distributions, with
densities and $\as$ evaluated at each corresponding order
(i.e., we use $(n+1)$-loop $\as$ at N$^n$LO, with $n=0,1,2$),
and we consider $N_f=5$ massless quark flavors. The default
renormalization ($\mu_R$) and factorization ($\mu_F$) scales are set to
$\mu_R=\mu_F=m_Z$.

The corresponding LO, NLO and NNLO cross sections as a function of $\sqrt{s}$ are reported in Fig.~1.
For comparison, we also show the NLO result supplemented with the loop-induced gluon fusion contribution (``NLO+gg'') computed with NNLO PDFs.
The lower panel in Fig.~1 shows
the NNLO and NLO+gg predictions normalized to the NLO result.
The NLO corrections increase the LO result by about $45\%$. The impact of NNLO corrections with respect to the NLO result ranges
from 11\% ($\sqrt{s}=7$ TeV) to 17\% ($\sqrt{s}=14$ TeV). Using NNLO PDFs throughout, the gluon fusion contribution provides between $58\%$ and $62\%$ of the full NNLO correction.
We find that the one-loop diagrams involving a top quark provide a contribution which is only few {\em per mille} of the full NNLO cross section. Since the quantitative impact of the two-loop diagrams with a light fermion loop is extremely small, we estimate that the neglected
two-loop diagrams involving
a top-quark contribute well below the {\em per mille} level.

The theoretical predictions can be compared to the ATLAS and CMS measurements \cite{Aad:2012awa,Chatrchyan:2012sga,ATLAS-CONF-2013-020,CMS-PAS-SMP-13-005} carried out at $\sqrt{s}=7$ TeV and $\sqrt{s}=8$ TeV, which are also shown in the plot.
We see that the experimental uncertainties are still relatively large and that the ATLAS and CMS results are compatible with both the NLO and NNLO predictions. The only exception is the ATLAS measurement at
$\sqrt{s}=8$ TeV \cite{ATLAS-CONF-2013-020}, which seems to prefer a lower cross section.
The comparison between our predictions and the experimental results, however, should be
interpreted with care. First, we point out that the LHC experiments obtain their $ZZ$ production cross section from four-lepton production using an interval in dilepton invariant masses around the $Z$ boson mass, thus not including some contribution from far off-shell $Z$ bosons. Then, EW corrections are not included in our calculation, and are expected to provide a negative contribution to the inclusive cross section \cite{Bierweiler:2013dja}.


\renewcommand{\baselinestretch}{1.6}
\begin{table}[ht]
\begin{center}
\begin{tabular}{|c| c| c| c|}
\hline
$\sqrt{s}$ (TeV) & $\sigma_{LO}$ (pb) & $\sigma_{NLO}$ (pb) & $\sigma_{NNLO}$ (pb)\\ [0.5ex]
\hline
7 & 4.167$^{+0.7\%}_{-1.6\%}$ & 6.044$^{+2.8\%}_{-2.2\%}$ & 6.735$^{+2.9\%}_{-2.3\%}$ \\
\hline
8 & 5.060$^{+1.6\%}_{-2.7\%}$ & 7.369$^{+2.8\%}_{-2.3\%}$ & 8.284$^{+3.0\%}_{-2.3\%}$ \\
\hline
9 & 5.981$^{+2.4\%}_{-3.5\%}$ & 8.735$^{+2.9\%}_{-2.3\%}$ & 9.931$^{+3.1\%}_{-2.4\%}$ \\
\hline
10 & 6.927$^{+3.1\%}_{-4.3\%}$ & 10.14$^{+2.9\%}_{-2.3\%}$ & 11.60$^{+3.2\%}_{-2.4\%}$ \\
\hline
11 & 7.895$^{+3.8\%}_{-5.0\%}$ & 11.57$^{+3.0\%}_{-2.4\%}$ & 13.34$^{+3.2\%}_{-2.4\%}$ \\
\hline
12 & 8.882$^{+4.3\%}_{-5.6\%}$ & 13.03$^{+3.0\%}_{-2.4\%}$ & 15.10$^{+3.2\%}_{-2.4\%}$ \\
\hline
13 & 9.887$^{+4.9\%}_{-6.1\%}$ & 14.51$^{+3.0\%}_{-2.4\%}$ & 16.91$^{+3.2\%}_{-2.4\%}$ \\
\hline
14 & 10.91$^{+5.4\%}_{-6.7\%}$ & 16.01$^{+3.0\%}_{-2.4\%}$ & 18.77$^{+3.2\%}_{-2.4\%}$ \\
\hline
\end{tabular}
\end{center}
\label{table1}
\renewcommand{\baselinestretch}{1.0}
\caption{Inclusive cross section for $ZZ$ production at the LHC at LO, NLO and NNLO with $\mu_F=\mu_R=m_Z$. The uncertainties are obtained by varying the renormalization and factorization scales
in the range $0.5 m_Z<\mu_R,\mu_F<2m_Z$ with the constraint $0.5<\mu_F/\mu_R<2$.}
\end{table}

In Table~1 we report the LO, NLO and NNLO cross sections and scale uncertainties,
evaluated by varying $\mu_R$ and $\mu_F$ simultaneously and independently
in the range $0.5 m_Z<\mu_R,\mu_F<2m_Z$ with the constraint $0.5<\mu_F/\mu_R<2$.
From Table 1 we see that the scale uncertainties are about $\pm 3\%$ at NLO and remain of the same order at NNLO. We also see that the NLO scale uncertainty does not cover the NNLO effect. This is not unexpected since the gluon fusion
channel, which provides a rather large contribution, opens up only at NNLO.

We have reported 
the first calculation of the inclusive cross section for the production of on-shell $Z$-boson pairs at the LHC
up to NNLO in QCD perturbation theory. The NNLO corrections increase the NLO result by an amount
varying from 11\% to 17\% as $\sqrt{s}$ ranges from 7 to 14 TeV. The loop-induced gluon fusion contribution
provides more than half of the complete NNLO effect.
Our calculation of the total cross section is based on the two-loop matrix element for $q{\bar q}\to ZZ$ for on-shell $Z$ bosons.  A computation of the two-loop helicity amplitudes will open up a spectrum of more detailed phenomenological studies at NNLO, including off-shell effects, differential distributions of the $Z$ boson decay products and direct comparison with the experimentally measured fiducial cross sections.

\noindent {\bf Acknowledgements.}
We would like to thank A.~Denner, S.~Dittmaier and L.~Hofer for providing us with the {\sc Collier}
library.
This research was supported in part by the Swiss National Science Foundation (SNF) under contracts CRSII2-141847, 200021-144352, 200020-149517, PP00P2-128552
and by 
the Research Executive Agency (REA) of the European Union under the Grant Agreements PITN--GA---2010-264564 ({\it LHCPhenoNet}), PITN--GA--2012--316704 ({\it HiggsTools}), and the ERC Advanced Grant {\it MC@NNLO} (340983).

\end{document}